\documentclass[a4paper,twoside]{article}

\usepackage{epsfig}
\usepackage{subcaption}
\usepackage{calc}
\usepackage{amssymb}
\usepackage{amstext}
\usepackage{amsmath}
\usepackage{amsthm}
\usepackage{multicol}
\usepackage{pslatex}
\usepackage{apalike}
\usepackage[bottom]{footmisc}
\usepackage{SCITEPRESS}     

\usepackage{natbib} 

\begin{document}

\title{Digital Twins for Trust Building in Autonomous Drones through Dynamic Safety Evaluation}

\author{\authorname{Danish Iqbal\sup{1}, Barbora Buhnova\sup{1} and Emilia Cioroaica\sup{2}}
\affiliation{\sup{1}Masaryk University, Brno, Czech Republic}
\affiliation{\sup{2}Fraunhofer IESE, Kaiserslautern, Germany}
\email{\{danish, buhnova\}@mail.muni.cz, emilia.cioroaica@iese.fraunhofer.de}
}


\abstract{The adoption process of innovative software-intensive technologies leverages complex trust concerns in different forms and shapes. Perceived safety plays a fundamental role in technology adoption, being especially crucial in the case of those innovative software-driven technologies characterized by a high degree of dynamism and unpredictability, like collaborating autonomous systems. These systems need to synchronize their maneuvers in order to collaboratively engage in reactions to unpredictable incoming hazardous situations. That is however only possible in the presence of mutual trust.
In this paper, we propose an approach for machine-to-machine dynamic trust assessment for collaborating autonomous systems that supports trust-building based on the concept of dynamic safety assurance within the collaborative process among the software-intensive autonomous systems. In our approach, we leverage the concept of digital twins which are abstract models fed with real-time data used in the run-time dynamic exchange of information. The information exchange is performed through the execution of specialized models that  embed the necessary safety properties. More particularly, we examine the possible role of the Digital Twins in machine-to-machine trust building and present their design in supporting dynamic trust assessment of autonomous drones. Ultimately, we present a proof of concept of direct and indirect trust assessment by employing the Digital Twin in a use case involving two autonomous collaborating drones.}

\onecolumn \maketitle \normalsize \setcounter{footnote}{0} \vfill

\section{\uppercase{Introduction}}
\label{sec:introduction}

The rapid adoption of new technological advancements within the automotive sector is causing major security concerns with direct safety implications. Consequently, approaches for co-analysis of safety and security concerns are emerging~\citep{safety_security_comparison, 9474627} with a particular focus on assuring the safety of autonomous systems subject to communication errors caused by malicious attackers.
A significant technological transition towards improved logistic behavior is the provision of autonomous drones that have the potential to enhance efficiency and optimized performance in various industrial applications~\citep{utne2020towards}. Autonomous drones are anticipated to advance conventional transportation mechanisms  by bringing significant  cost reduction, improved safety and positive environmental impact.

With the growing risk of malicious attacks, however, the safety of the operational environment and of the systems needs to be ensured. In the specific context of trust, safety is an essential property~\citep{avivzienis2004dependability}. In our scenario, the updates of drones with complex software behavior and advanced sensors, actuators and advanced capabilities to communicate with other agents within the ecosystem, the complexity of the operational domain increases considerably. This aspect is of extended concern in situations when the increased autonomy of collaborating autonomous systems relies on a trustworthy software-hardware interaction. The complexity that the dynamic runtime updates bring within a smart digital ecosystem highlights new challenges within the safety community. 


In general, together with the  deployment of innovative technology within autonomous ecosystems, the quality of information exchange plays an essential role~\citep{tao2018digital}. Namely, information can be exchanged between two collaborating systems to support trust building between them~\citep{IqbalSMC2022}, when one system can share its planned behaviour with its collaborator in an attempt to disclose information about it and foster trust building. 
Digital Twin (DT), which is a virtual model reflecting physical objects within the digital world, can become such information, used in the exchange to support the trust-building process. In our case, this information needs to include the safety-relevant properties that can be dynamically evaluated at runtime.  
The idea of employing DT of an autonomous agent in building trust with its collaborator is very promising. So far, it has only been explored in~\citep{IqbalSMC2022}, where it is however unclear what information shall be included in the actual Digital Twin, and how different safety rules could be reflected in it. 
 
Dynamic safety analysis of collaborating autonomous systems or agents within the autonomous ecosystems is a challenging task for both the industry and academia~\citep{asaadi2020dynamic}. To avoid runtime exploitation of vulnerabilities of critical systems, runtime dynamic mechanisms that can perform a safety assessment for concrete technical situations is needed~\citep{leng2021digital}. The dynamic open context of operation for autonomous systems implies that it is constantly evolving, making it hard to anticipate how safe an autonomous system will be in that context. For autonomous systems, in particular, it becomes challenging to dynamically evaluate the predicted behavior of the autonomous agent, which is why the DT provided by the collaborator (trustee) can be of enormous help. The trustor can then employ the DT of the trustee in runtime compliance checking of the actual trustee behaviour with its declared behaviour (delivered via the DT)~\citep{cioroaica2022paradigm}. This is a concept we elaborate in this work, with a focus on the incorporation of relevant safety rules in the DT specification. 

Consequently, in this paper, we present a method for trust evaluation between collaborating autonomous agents, used in the context of autonomous drones, which employs the concept of a Digital Twin (DT) enriched with safety properties. The DT can be exchanged between  collaborating agents with the aim of building trust at runtime. By applying the STPA (Systems-Theoretic Process Analysis~\citep{dghaym2021stpa}) rules, we present a concept for dynamic safety evaluation used for building trust between collaborating autonomous drones. Moreover, our dynamic safety evaluation approach aims to improve risk assessment accuracy by employing compliance checking for safe-behavior assessment. For this, runtime behavior intentions are turned into virtual trajectory distributions and are virtually analyzed  using behaviour-specific framework for hazard analysis. 
Our proposed hazard analysis framework supports the addition of safety attributes to the DT specification. Finally, we present a proof of concept to illustrate the method on an example of collaborating autonomous drones.

The paper is structured as follows. After the introduction in Section~\ref{sec:introduction}, Section 2 describes the state of the art, Section 3 illustrates our trust-building method with safety assurance rules, Section 4 describes the evaluation and Section 5 concludes the paper. 

\section{\uppercase{State of the Art}}

Safety is a highly desirable property that requires assurance during the construction of systems that operate within safety-critical autonomous ecosystems~\citep{widen2022autonomous}. The typical approach employed in the process of safety analysis of autonomous safety-critical systems relies on the identification of the causes of the system's potential risks and uncertainties~\citep{tauber2018enabling}. Then, based on the analysis of these risks and uncertainties, measures to minimize the critical system's failure chances  are implemented~\citep{koopman2021ethics}. In this context, this section discusses the role of Digital Twins (DT) in connection to the process of assuring safety and building trust in autonomous safety-critical systems.

\subsection{Trust Building via Employing Digital Twins}
The increasing uptake of autonomous drones, robots, and similar AI-enhanced autonomous systems within human societies requires an increased evaluation of runtime trustworthiness of safe operation~\citep{hancock2011can,wagner2009role}. The significance of trust has been highlighted in the human-robots interaction~\citep{kuipers2018can}, and has been widely researched in psychology, management, economics and philosophy~\citep{lahijanian2016social,huang2019reasoning}. An important aspect in the process of building trust in autonomous systems needs to account for the dynamic safety evaluation of predicted behavior. This can be done with the support of run-time verification mechanisms~\citep{vierhauser2019interlocking}, possibly with the help of Digital Twins (DTs) used as a means of information exchange. In this paper, we propose the employment of DT to integrate the safety rules and mechanism to provide needed safety information to a collaborator in order to support trust building among autonomous agents, following the concept outlined in our previous work~\citep{IqbalSMC2022}. 

Within various industries, IoT (Internet Of Things) technologies are developed to meet the needs of diverse systems such as robots, and intelligent devices~\citep{kochovski2018supporting}. In order to keep up with the runtime needs of a system operation, the National Aeronautics and Space Administration (NASA) and the United States Air Force (USAF) were the first to adopt the concept of DT used to predict the life span and maintenance of a spacecraft~\citep{negri2017review}. The application of a DT then became a popular approach used for synchronizing the physical world with the digital world~\citep{liu2020framework,qi2021enabling}. 
Moreover, the interaction between the physical world and the virtual digital world enables systems in various industries to achieve high levels of automation and intelligence~\citep{boje2020towards}. Many studies and researchers employ the DT in various industrial sectors, such as Industry 4.0 and production with the emergent potential of DT runtime execution  for building trust~\citep{cioroaica2019not}.

\subsection{Safety Analysis}
Traditionally Leveson’s Systems-Theoretic Accident Model and Processes (STAMP)~\citep{leveson2018stpa} has been used as an accident model based on system theory, improving reliability and supporting conventional analytical reduction of risks. The model primarily starts from the premise that accidents occur as a consequence of complex, dynamic inter-related processes rather than as a result of a series of component failures. 

Then, the systems-theoretic process analysis (STPA)~\citep{dghaym2021stpa} is a hazard analysis method that incorporates the fundamental concepts of the STAMP accident causality models. STPA is a top-down system engineering approach to assuring system's safety that can be employed in the early phases of  system development with the scope of creating  high-level safety requirements and constraints~\citep{plioutsias2018hazard}. However, the approach has limitations on  supporting the generation of a probability value associated to a hazard. This happens because the strategy to compute an accident probability requires the exclusion of critical causal factors that are not stochastic or do not have probabilistic data. Therefore, STPA aims to investigate unsafe control actions by identifying the path to accidents or risk scenarios~\citep{ozerovsafety}. Then, it incorporates causal factors that traditional analysis techniques do not fully consider, such as: uncontrolled component interactions, incorrect requirements, insufficient coordination between multiple controllers, and poor decision management~\citep{plioutsias2018hazard}. As a result, unsafe behavior is treated as a dynamic control problem rather than a component's reliability problem.

   \begin{figure*}[!htbp]
  \centering
  \includegraphics[scale=0.42]{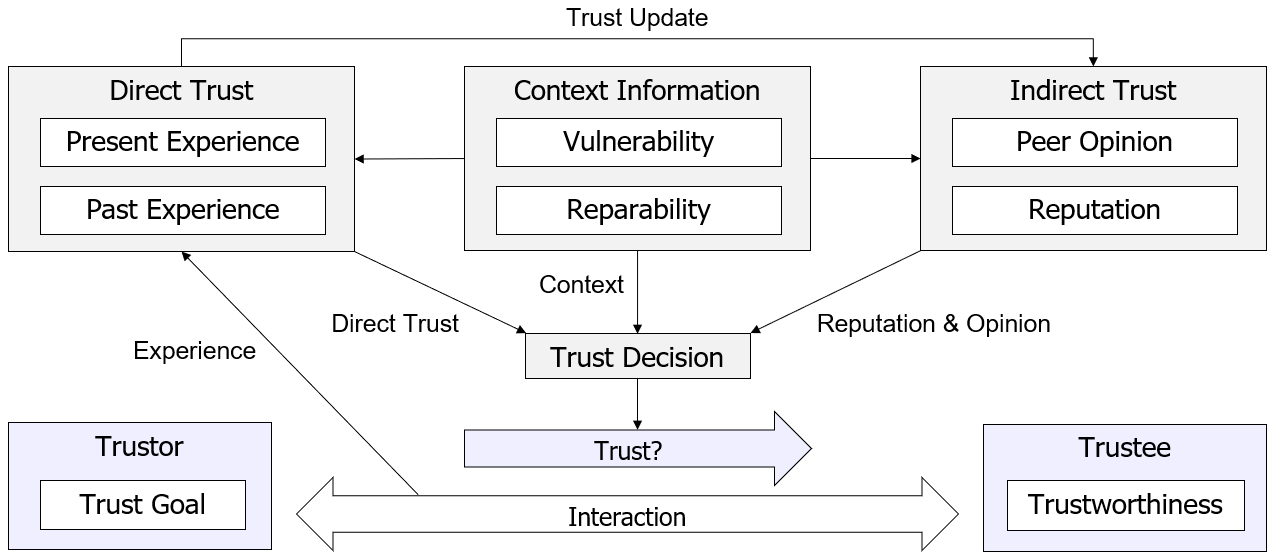}
  \caption{Direct and Indirect Trust~\citep{buhnova2022tutorial}}
  \label{fig:trust-management-components}
 \end{figure*}

\section{\uppercase{Safety-Based Trust-Building Method}}

By elevating from the traditional safety assurance processes, in this paper, we propose the concept of dynamic safety assurance used for building trust between two autonomous agents based on run-time compliance checking of trustee actual behaviour against it declared behaviour provided in form of its Digital Twin. To achieve this goal, we propose a safety assurance process embedded within the Digital Twin by incorporating system's properties that emerge from applying the STPA process directed towards dynamic safety evaluation. The proposed safety method includes five steps, which are explained in detail below. 

The notion of trust relies on how effectively two agents collaborate within an ecosystem. The collaboration among autonomous agents can be divided into two categories as direct and indirect trust~\citep{buhnova2022tutorial,byabazaire2020data}. The combination of these trust-building elements can be explained further with the help of Figure~\ref{fig:trust-management-components}, which exemplifies how direct and indirect trust contribute to the trust decision of a trustor in relation to a trustee. Direct trust can be built through the direct experience of the trustor with the trustee and results from multiple interactions. In our approach, the experience is gained via the compliance checking of the trustee behaviour with its declared behaviour incorporated within the DT that it shares with the trustor. This is done to reflect two major social metrics relevant to trust building, which are: 1) \emph{honesty} (consistency between the declared and actual behaviour) and 2) \emph{openness} (transparency about the intended behaviour). On the other hand, the indirect trust manifests the opinions of the neighbors (peer network) and the gathered reputation of the trustee from previous entities that have engaged in collaborations with the trustee. We further detail the role of the direct and indirect trust below.

For the definition of safe behavior, we have followed the STPA process for deriving the unsafe action control and identifying the causal factors, which are categorized further in the safety assurance process.
Below we are detailing the steps in creating DTs that can be used for the dynamic safety evaluation with the Systems-Theoretic Accident Model, Processes Model and Systems-Theoretic Process Analysis Technique.

\subsection{Direct Trust via Run-Time Compliance Checking}
To assess the direct trust in agent-to-agent interaction, we suggest to employ run-time compliance checking among the collaborating agents with the help of the Digital Twins, representing the declared behaviour disclosed in the first phases of the trust building process.
The direct trust shows the direct experience of the trustor with the targeted autonomous agent. Moreover, the direct trust of trustor agent in the trustee agent is built through direct collaboration and run-time compliance testing using the DT of the trustee to check the consistency in their declared and actual behaviour. If the trustee's actual behavior matches the declared behavior shared in the form of the DT while taking the safety requirements into consideration, the trust experience of the trustor reflects it into an increased willingness to trust the trustee.

\subsection{Derivation of Safety Rules via the STPA Method}

To assess the hazards and the associated factors in the operation of a drone system, this section discussed how to determine individual safety requirements. The individual steps of the STPA process are discussed below as a mechanism to specify the system under study and the safety method scope.


STPA~\citep{leveson2018stpa, dghaym2021stpa} is a method for hazard detection that uses control structures to represent the system behavior. The hazardous conditions are identified through the unwanted presence (commission), absence (omission), or improper timing (delay/wrong value) of control actions. Then the causal factors for unsafe control actions are identified. by inter-playing formal specifications in the verification process, the systems' reliability can be assured. Approaches that integrate formal method with analysis techniques such as STPA offer greater support for trust building in the system based on safe and correct behavior that results in a reliable runtime execution. Such a process can be divided into three phases described below.

\medskip
\noindent\textbf{Phase 1: Hazard Identification and Analysis}

\begin{figure*}[!htbp]
  \centering
  \includegraphics[scale=0.55]{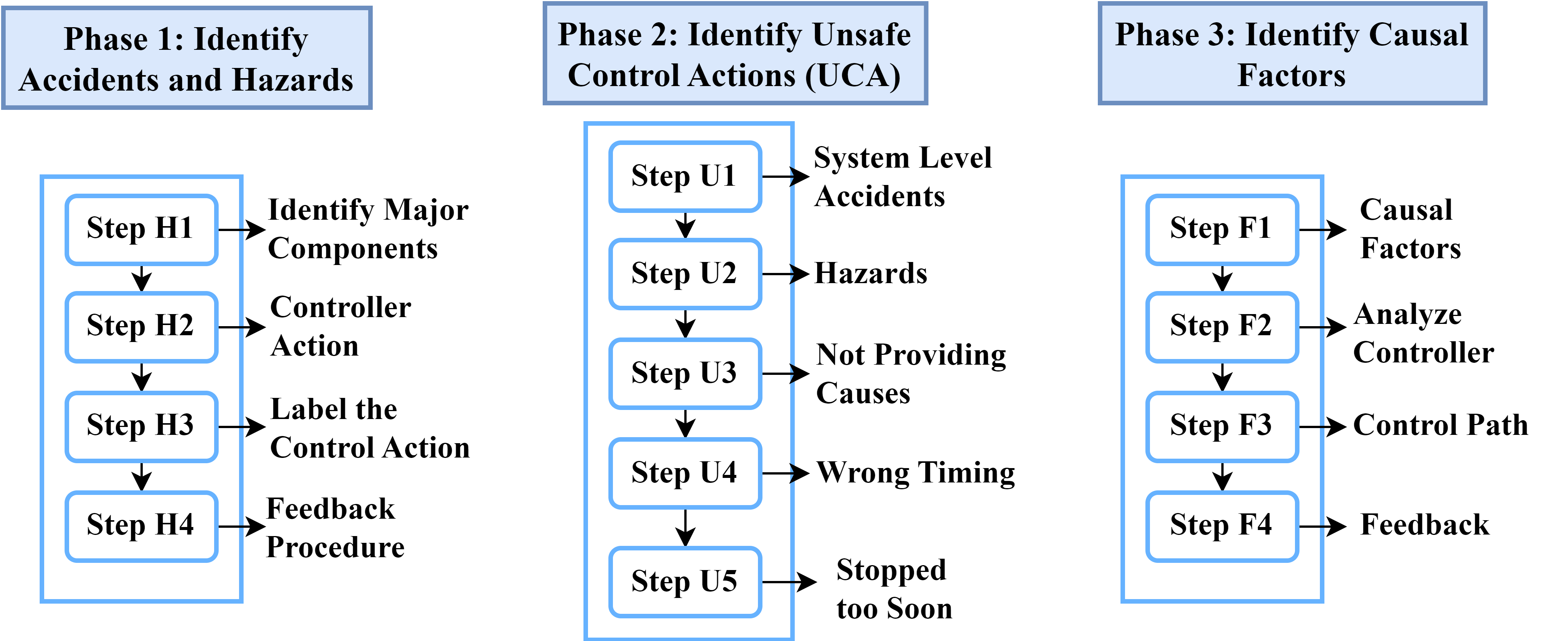}
  \caption{Framework for Identification of Hazard, Unsafe Control Action and Casual Factors}
  \label{Fig:Framework}
\end{figure*}

\noindent In this phase, system-level hazardous events and major risks of accidents are analyzed and identified. The system-level safety limits are defined at the interface of interaction with other components outside the scope of the system.  For our use case, we define how unsafe control actions (UCA) and scenarios in which the UCA may occur may violate system-level safety restrictions and cause a transition into unsafe state that can cause hazardous situations.
The final step is to identify the triggers of these  hazardous situations with greater chance of occurrence. These situations act as preconditions, facilitators, or enablers. Figure~\ref{Fig:Framework} depicts the essential phases of identification of hazardous situations and unsafe control action,  which are listed below.

\begin{itemize}
 \item \textbf{Step H1:} Describe the system as an independent control system by giving it a definition, including its components and boundaries. For example, according to hazards and accidents, the major components of an autonomous drone system are the physical drone, software controller and navigation system. The larger context of use within dynamic autonomous ecosystems account for the existence of additional entities such as an ecosystem orchestrator, collaborating drones and ecosystem actors (businesses, field workers, etc.).
 \item \textbf{Step H2:} Define system-level hazardous events and safety constraints. This is refined into the  identification of the controller responsibilities and process variables.
 \item \textbf{Step H3:} Elaborate the unsafe control actions (UCA) that violate the safety requirements and label them according to the STPA process.
 \item \textbf{Step H4:} Define the procedure for feedback when a hazard or accident occurs.
\end{itemize}

\smallskip
\noindent\textbf{Phase 2: Identification of Unsafe Control Actions} 

\noindent The unsafe control action in Phase 2 is described as follows:
\begin{itemize}
\item  \textbf{Step U1:} The system-level problem is defined. An accident example it is the crash of the physical drone system. How 
the overall system (physical drone, software controller, and navigation system) or ecosystems (formed by the orchestrator who implements the logistic algorithm together with the communicating drones) responds to unsafe control actions and accidents?
\item  \textbf{Step U2:} Hazard is defined in the unsafe control action as the Drone under evaluation is violating the minimum separation distance to another drone, and what are the external control possibilities, when the system fails to adhere to a safe behavior? In case of internal self-adaptation mechanisms, how does the system respond and reconfigure itself in case of an hazardous situation?
\item  \textbf{Step U3:} When the hazard or accident is not linked to a cause, then an external controlling entity (that can be an ecosystem orchestrator that implements intelligent logistic algorithms for recovery) needs to create a plan of action that involves cooperation of the remaining drones and actors within the ecosystem (such actors can be businesses, but also engineers, and field workers, etc.). In case of autonomous drone behavior, the system needs to reconfigure and self-adapt itself internally.
\item  \textbf{Step U4:} During an hazardous situation that manifests in an accident, an important aspect of a recovery procedure is the time management that reflects by the capacity of the ecosystem, or of the system itself to tackle the problem within a specified time frame.
\item  \textbf{Step U5:} As a general rule, the monitoring of the failed system shall not stop too early during the violation of the rules or during the unsafe control action. Sometimes the system can come to track in the ending phase.
\end{itemize}

\smallskip
\noindent\textbf{Phase 3: Identification of Causal Factors}

\noindent This phase is concerned with the identification of causal factors, and is explained as follows:
\begin{itemize}
 \item  \textbf{Step F1:} Definition of the causes of hazards that can lead to accidents, such as connection problems, environment negative influence, or triggering of internal faults that manifest into system failure. Under these situations, the system action need to be defined.
 \item  \textbf{Step F2:} Definition of factors and risks to be analyzed by the external controller (ecosystem orchestrator) or internal controller (in case of autonomous behavior) and definition of a plan of action to be communicated within the ecosystem. This plan of action needs to be shared  with other ecosystem components (such as drones, ecosystem actors that can intervene in case of an accident).  
 \item  \textbf{Step F3:} Develop a recovery path for the ecosystem according to the different hazardous situations.
 \item  \textbf{Step F4:} Develop a feedback mechanism that accounts for the risk factors and correlated hazardous situations. Such a feedback mechanism supports the improvement of the recovery strategy.
\end{itemize}

\subsection{Derivation of Safety Properties for Dynamic Run-time Evaluation}
We start from the definition of the Digital Twins to be abstract models fed with real-time data~\citep{cioroaica2022paradigm}. For the creation of digital twins used for the dynamic evaluation of a system's safety, the safety properties need to be selected from the overall behavior specification. This is performed following a standardized safety procedures, such as the STPA / Standard Safety procedure for Functional Safety.

\begin{figure*}[!htbp]
  \centering
  \includegraphics[scale=0.038]{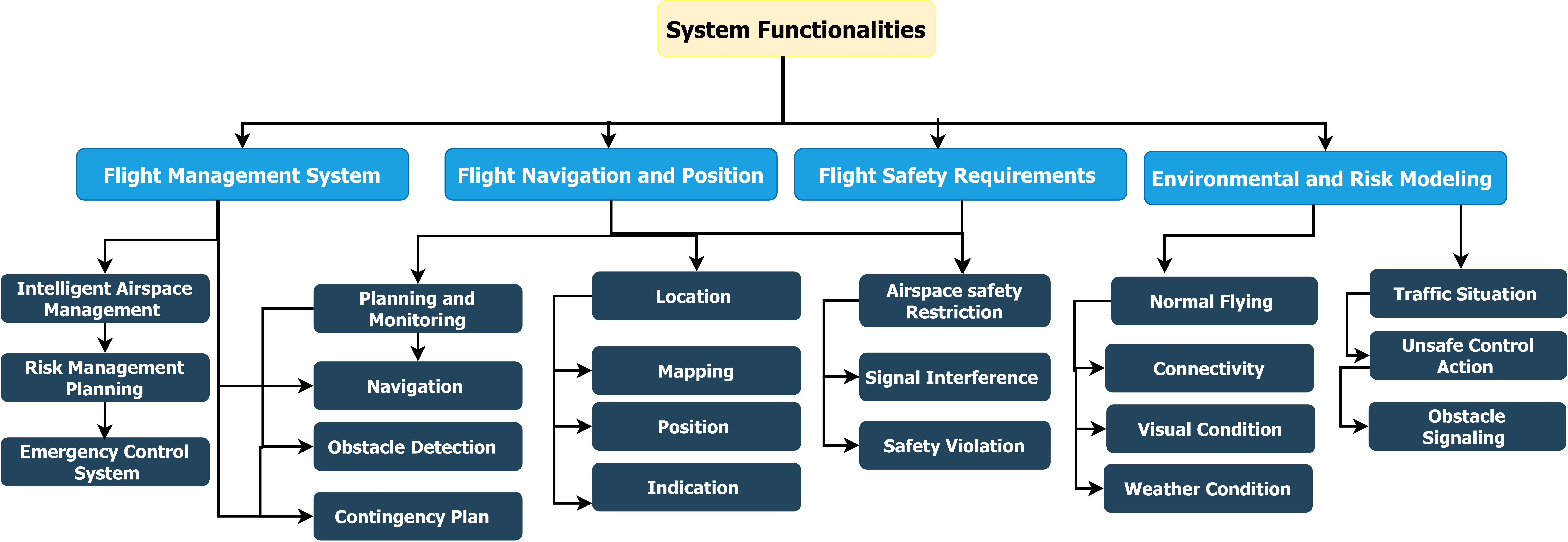}
  \caption{Drone System Function }
  \label{Fig:Drone-prop}
\end{figure*}

The expected autonomous drone behavior relies on the drone's  capabilities to execute core functions (nominal and operational fail-over behavior) to perform its flying safely and efficiently. Figure~\ref{Fig:Drone-prop} depicts the various functions of the system, which are described as follows.

\textbf{Flight Management System:} 
The Drone is controlled during flight using an intelligent mission airspace management system. This system shifts the human role of drone control through a specialized autonomous controller. In order to ensure a safe  behavior,  and achieve a high-level safety goal, it implements an internal dynamic risk management and emergency control loop. To accomplish such goals, the drone's managing control system uses monitoring of the controller specialized in autonomous flying (the one that replaces human control) and feedback mechanisms to regulate communication between the sensors, actuators and the drone's other subsystems that can be triggered in case of detected deviations. 

\textbf{Flight Navigation and Position System:} 
A number of functions is needed to perform way-point navigation and position of the Drone from the source to the destination. First, flight planning and environment monitoring are important components of a flight navigation system. Then, a component for obstacle detection is a key element on which the safe behavior within the navigation phase relies. A fast detection of obstacles can minimize the risk of hazardous situation. Third, the system's reaction plan and its capability internally adapt to unforeseen events needs to be considered.

\textbf{Flight Safety Supervisor:}  
A flight safety supervisor is required in order to execute the flying mission safely and make decisions at the right time. The ecosystem's airspace safety restriction can protect the drones from collisions. In context of multiple drones that operate in an open environment together with other systems, avoiding signal interference and physical safety violation distances are important for a safe synchronized flight.

\textbf{Environmental and Risk Monitoring:} 
The Environment and risk monitoring component is essential to execute the mission safely and avoid drone accident. The timely reaction of the system relies on the proper functionality of this component. Other role, this component has is in minimizing the risk it imposes to the environment, under special operating conditions such as weather change. Moreover, the management of an internal risk model and action plan according to traffic conditions, unsafe control actions and obstacle signaling needs to be implemented in order to ensure a safe flight missions. 
    
\subsection{Derivation of Internal Behavioral Transitions} 
Building trust in the autonomous drones over time is facilitated by modeling physical information in the form of a DT based on real-world system behavior. With the help of run-time compliance checking, the Digital Twins simulate the physical state of its corresponding autonomous drone. Recently, we have in~\citep{IqbalSMC2022} shown that modeling the interaction behaviour in form of a Digital Twin can be a very promising way to establish trust between collaborating drones. In this paper, we propose to build the DT on the basis of safety rules and considerations discussed above, and present an example of such DT model in the next section. During the creation of the DT, which can have e.g. form of a Finite-State Machine (FSM), we consider the safety properties of an autonomous Drone as the major aspects to be included. The safety properties are included according to the STPA safety rules for the unsafe action control. Finally, the design of the DT along with the safety properties is explained in an example below (and depicted in form of an FSM in Figure~\ref{Fig:Digital-twin}). 


\subsection{Indirect Trust via Reputation-based Derivation of Expected Cooperation}
Safe collaboration between two drones can be reported to contribute to the reputation-building of the drones, and creation of direct and indirect trust. The indirect trust can reflect the past experience of the trusted neighbors and other nearby autonomous drones that engaged in previous collaborations with the target autonomous drone (the trustee). Some researchers have previously associated experience and reputation to explain indirect observation from a reputable authority. As discussed above, in case an ecosystem has an orchestrator, the orchestrator can become the central authority that provides the evidence and claim of trust for the trustee.  on the other side, the direct trust is navigated through a direct link between the trustor and the trustee relying on the trustor  confidence that the trustee will carry out the expected task. In this case, reputation aggregates all the prior experiences with the trustee. 



\section{\uppercase{Evaluation}}
This section presents a case study of developing trust based on dynamic safety evaluation among autonomous drones employed in the logistic domain. To this end, we extract the case study components, walk through the process over the use case, and present the properties of the autonomous drone based on safety regulation within the specific scenario. Then, we illustrate the representation of these properties in the form of an FSM used in the construction of the Digital Twin (DT) of an autonomous drone. Finally, based on the DT and run-time compliance checking with the actual drone behavior, we evaluate the behavior of the drone, advising whether or not it can be trusted.

\subsection{Use Case}
\citet{hussein2021key} introduced a case study within the logistics domain directed towards performing a simultaneous delivery to multiple locations at a feasible price and in a timely manner. The use case addresses the challenges of the access in remote areas that do not have transportation infrastructure. The use case generates two demonstrations to showcase the capabilities of autonomous drones.

The first demonstrator objective is to deliver geophysical sensors by using a fleet of autonomous drones, as an autonomous Drone can reach hard-to-access locations. The Drone builds an innovative 3D high-density geophysics solution that improves the caliber of subsurface images and reduces costs and environmental imprint. 

The second demonstrator goal is to deliver a package in a hospital using a droid (to carry it inside a building) and a drone (to carry it between buildings) to accomplish fast delivery. The demonstration is accomplished in two stages. In stage one, which is the longer one, the design of the droid and the Drone is done in an industrial site to  demonstrate the capability of their coupling to transport light parts. The second stage intends to replicate the industrial site experiment in hospitals for transporting medicine or test samples.

\subsection{Properties Extraction} 

Iteration of the process over the use case is directed towards providing evidence of validity via the steps presented above. In the first step, we present the scenario and the use case. In the second step, we extract the properties of the Drone use case by the safety rules defined in Section 3 of the method. Then, following the safety process, we extract and incorporate the safety rules within drones' explicit behavior. The third step involves modeling these attributes and integrating the information that is presented in the form of an FSM model representing the Digital Twin. In the end, we explain the scenario within an example that supports the trust decision.

\subsection{Autonomous Drone Safety Requirements}
In this section, the properties of an autonomous drone are extracted from the use case, following the safety process in the method in Section 3. These properties are based on the work of~\citet{plioutsias2018hazard}, following the drones safety analysis process.

\medskip
\noindent \textbf{Design-time Safety Requirements:}
The safety requirements for an autonomous Drone in our approach consist of the following: 
\begin{itemize}
\item The Drone system complies with civil aviation norms, meaning that only drones that weigh less than 20 kg are taken into consideration. This aspect is decided in the early phases of drone design. A property of "weight" can be explicitly communicated at runtime to a collaborator. 
\item The physical drone interaction with other components such as sensor and flight controller were analyzed with respect to items mentioned above. 
\item The civil aviation authority's airspace safety regulations are taken into consideration.
\item The civil aviation authority's license requirements and airworthiness standards are taken into consideration.
\end{itemize}

\smallskip
\noindent \textbf{Flight Management System:}
The safety properties of the flight management and control system  for autonomous Drone are described below.
\begin{itemize}
\item The response to critical situations of the emergency control system is considered in terms of time property (too late, too early, non existent).
\item Connectivity to the network with the control tower is considered with respect to status (on/off) and up/down time.
\item Operation regulations for small drones are taken into the consideration with respect to speed properties.
\item Incident and accident report is considered for specified period of up time (time of flying).
\item Rules for external communication are followed according to the safety regulation and quantified as a ration: number of established communication/number of meetings.
\end{itemize}

\smallskip
\noindent \textbf{Flight Navigation and Position (Collision Avoidance):}
The following is a description of the safety requirements and emergent safety properties for collision avoidance to aid in the development of trust among autonomous drones.
\begin{itemize}
\item The eventuality of collision with other flying drones has been computed by involving the risk management scheme and has been quantified.
\item The Drone keeps updating the neighboring Drones about the information of turning right or left and increasing or decreasing the speed.
\item The Drone establishes a minimum separation of 40 meters from other drones.
\item The Drone implements check run-time signal interference. 
\end{itemize}

\smallskip
\noindent \textbf{Environmental and Risk Modeling:}
The following is a description of the environmental and risk modeling to execute mission safely flight and analyze the environmental risk.
\begin{itemize}
\item The Drone implements the unsafe action control to avoid external disturbance caused by weather change.
\item In bad weather conditions, the Drone increases distance in order to maintain visibility and avoid hazardous situations.
\item The Drone continuously establishes connectivity with the central ecosystem orchestrator in order to receive help on avoid hazardous situations.
\end{itemize}

\subsection{FSM Model of Drone Digital Twin}
By integrating the engineering knowledge of the system's architecture and safety knowledge of nominal and failure behavior in the early stages of system development, state machines can be defined to support the DT. After modeling the behaviour, safety  information can be reflected in the FSM in the form of various states. Later on, as the specification of each system component advances, the high-level models are replaced with detailed sates of the FSM. The safety knowledge can be integrated using a machine-readable representation of system nominal and failure behavior.

\begin{figure}[b]
  \centering
  \includegraphics[scale=0.38]{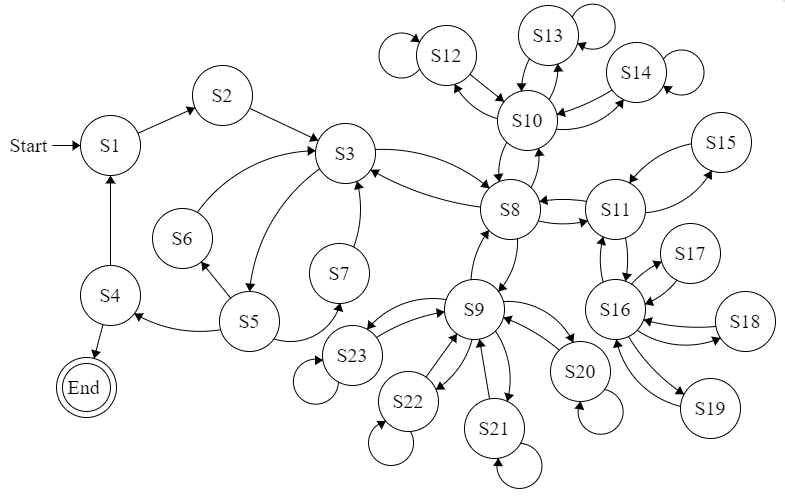}
  \caption{Digital Twin of an Autonomous Drone via FSM}
  \label{Fig:Digital-twin}
\end{figure}

\begin{table*}
	\renewcommand{\arraystretch}{1.2}\small
\caption{FSM States and Description}
\label{tab:definitions}
\centering
\begin{tabular}{|p{2.5cm}|p{13.5cm}|}
\hline
\textbf{State Name} & \textbf{Description} \\
\hline
S1 Accelerate to Start &  S1 is the start state of the Drone, which is considered as off at the time. When the Drone is turned on, its state changes from S1 to S2. \\
\hline
 S2 Air Trajectory Planner Updates & The S2 state deals with updates to the air trajectory planner, which tracks adjustments to the autonomous Drone's air trajectory. If Drone is moving forward, it will transit from S2 to S3. \\
\hline
S3 Flight Navigation Updates & The S3 state deals with flight navigation, and if an object is found, a transition to S5 is triggered; otherwise, the transition is triggered to S8 (Drone signal interface). \\
\hline
S4 Decelerate to Stop & S4 is the stop state of the Drone. As soon as the transition is complete, it goes to the end state.  Whenever an unexpected event occurs or an obstacle is detected, the Drone enters a stop state for a short time until the clear transition to S1 is completed. \\
\hline
S5 Object Detection & The S5 state deals with object detection. A transition to the S4 (stop state) is initiated if an object is detected and takes a long time to clear or if the Drone completes their transition. Unless the object is still detected, however, the transition is triggered to the S6 state (contingency plan) to avoid the obstacle. If an object is still present, the transition to S7 (lane changing) state takes place. \\
\hline
S6 Contingency Plan  & The S6 describes a contingency plan to prevent internal or external failure. If the action in the contingency plan is followed as expected, the Drone starts moving forward and enters the S3 state. \\
\hline
S7 Lane Changing & A lane changing is captured by S7. When a traffic obstruction is encountered, the Drone switches lanes; after changing lanes, when normal flow resumes, start moving forward, which triggers the transition to S3. \\
\hline
S8 Drone Signal Interface  & The S8 state describes the drone signal interface. The signal interface state has three options for deciding the Drone next transition. It is related to speed; therefore, the transition is triggered, resulting in the S9 state. If a lane change is required, the transition is initiated, resulting in the S10 state, or the Drone needs to stop so the transition is triggered, leading to the S11 state. When the Drone needs data for other flights, the transition is triggered to S3 (flight navigation updates).  \\
\hline
S9 Speed Information & The S9 state deals with speed information, and based on the situation, a transition to the states S20 (maintain distance), S21 (speed keeping), S22 (accelerate speed), S22 occurs (decelerate the speed), or the Drone returns to S8 (Drone signal interface).   \\
\hline
S10 Lane Keeping Information & The Drone expresses motion action that keeps the lane or changes to S12 (right turn), S13 (left turn), or S14 (level changing), or the Drone returns to S8 (Drone signal interface). \\
\hline
S11 Stop Information  & The S11 state represents the situation when the Drone is advised to stop (S16) without details of the reasons, and if the issue is analyzed as related to the environment, the transition is triggered to S15 (environmental issue) S15, or the Drone returns to S8 (Drone signal interface). \\
\hline
S12 Right Turn  & The S12 state represents the right turn. When the Drone makes a right turn in response to air traffic signal updates, follows the path, and then needs to make a turn again owing to traffic problems, the transition to the S10 state is triggered.  \\
\hline
S13 Left Turn  & The S13 state presents the left turn. The Drone changes the left turn according to the air traffic signal updates and follows the path, and then needs to make a turn again owing to traffic problems, the transition to the S10 state is triggered. \\
\hline
S14 Level Changing  & The S14 state deals with the level changing.  Depending on the path and other circumstances like the weather, the Drone adjusts its level. When the Drone needs to change the direction so the transition is triggered to the S10 state.  \\
\hline
S15 Environmental Issue  & The S15 state deals with environmental issues. The decision to discontinue was made because of environmental concerns. If the environmental problem is severe, the transition to S11 state (stop information) is triggered in accordance with the circumstances.  \\
\hline
S16 Brake & The S16 describes the brake state.  Breaking is needed, a decision about whether transition is triggered to S17(light brake) or S18 (strong break) or S19 (emergency brake) would happen according to the situation. Otherwise, the transition is triggered, to S11 (stop state). \\
\hline
S17 Light Brake  & The S17 state represents the Drone's light braking owing to traffic or other Drones coming in the route. In case of an obstacle, a transition is triggered to S16 (brake state). \\
\hline
S18 Strong Brake  & The S18 state represents the strong brake. Once the Drone reaches the stop position, and needs to change the state, a transition is triggered to S16 state. \\
\hline
S19 Emergency Brake & The S19 represents emergency brake, when unanticipated events happen that force the Drone to abruptly brake. The S19 state (key to emergency braking) triggers a transition to S16 according to the situation. \\
\hline
S20 Maintain the Distance & The S20 state deals with distance maintenance.  The S20 state is responsible for preventing Drone collisions and ensuring safety by keeping an appropriate distance between the two Drones. After maintaining the distance, a transition is triggered to S9 state.  \\
\hline
S21 Speed Keeping  & The S21 defines the speed keeping of the Drone. Until there is no conflict and the route is safe, the Drone maintains its current speed, otherwise the state changes to S9. \\
\hline
S22 Accelerate the Speed  & The motion is expressed by the S22 state. When there is no conflict and the route is safe and clear, the Drone accelerates, otherwise state changes to S9. \\
\hline
S23 Decelerate the Speed & The S23 expresses the control action. The Drone decelerates the speed based on the recognized situations such as traffic, hazards, objects, environments and states change information. Otherwise, a hazard is not present, the transition to S9 state is triggered and the forward motion is resumed. \\
\hline
\end{tabular}
\end{table*}

A simple example of FSM for a possible Digital Twin of an autonomous Drone is in Figure~\ref{Fig:Digital-twin}, together with the transitions between states that enable the expression of the behavioral properties and inherent safety properties. The contingency and collision avoidance actions included in the FSM are based on the work of~\citep{hejase2018identification}. 

After defining the state machine for a system, engineers identify the safety status of the system for each state. The safety status is related to the initial derivation of safety goals. Then, based on the state-machine definition, actions for every state are defined in order to assure the safety of the overall system operation. Consequently, every state then depicts a distinct operational context of the system by incorporating actions to reduce or eliminate unacceptable operation risks within the context for which the safety goals have been defined. The processing of the steps depends on the specifics of an application scenario. In a self-adaptive system, the actions become input events to other automated controls.

\subsection{Direct Trust via Run-Time Compliance Checking}
Based on the relevant hazard, emergency and background knowledge all the states and their transitions are added to the autonomous Drone Digital Twin through an FSM (see Figure~\ref{Fig:Digital-twin}). The aforementioned DT properties are verified against  the actual Drone behaviour via run-time compliance checking, where the actual compliance serves as an evidence to increase trust in the Drone. On the other hand, when the actual behaviour deviates from the DT behaviour, indicating a violation of the declared contract and breaking the trust, the Drone launches appropriate safety measures, e.g. avoiding the Drone, minimizing the speed or stopping. 

In our case, a collision risk between two autonomous drones can be reduced by maintaining the minimum distance and then maintaining the distance as realized in S20 in Figure~\ref{Fig:Digital-twin} in order to estimate the future locations of an autonomous drone given its current location, the way-points it is traveling to (the remaining mission), and the current speed and acceleration. As the Drone periodically compares the received locations with its own anticipated locations, Digital Twin broadcasts its future locations. 

 
\subsection{Indirect Trust via Reputation-based Derivation of Expected Cooperation}
Indirect trust is built over time based on third-party interactions with the trustee drone. The established reputation is a propagating property of trust. This form of trust is important as it informs the other drones about the past behaviour of the Drone in question. Moreover, reputation is also based on the quality of data assessment from the previous different components, such as sensors or other nodes. In the proposed scenario, the data is collected from different sources and intelligent data monitoring saves the data that the drones shared with each other in the from of DT. Later on, these data can be used as a third-party source with other agents, feeding the indirect trust. 


\section{\uppercase{Conclusion}}
This section discusses the key findings of the research to better comprehend the trust building in ecosystems using DT via dynamic safety evaluation. 

First, we studied how the concept of dynamic safety evaluation can support the building of trust in the dynamic ecosystems. With the integration of safety mechanism and run-time compliance checking, we evaluated the trust in the proposed scenario. We used the concept of a Digital Twin to capture the information from the physical objects. In the method, we discussed both direct and indirect trust. The direct trust can be obtained via run-time compliance checking of the declared (DT) and actual behaviour of the autonomous Drone, with the help of safety rules. The indirect trust can be obtained through the reputation, built form the data propagated from the third-party past interactions and feedback.

Second, we evaluated the proposed method by presenting a use case from the logistics domain where two drones interact with each other following the safety rules and using the DT for run-time compliance checking to build the trust in one another. We also discussed the direct and indirect trust on the practical use case, with the direct trust based on the proposed properties of the autonomous Drone and considering the safety rules to design the DT of the autonomous Drone in the form of an FSM. Finally, we presented the trust/not trust example for collision avoidance. 

In the future, we plan to improve the current design of the Digital Twin by including more detailed behavior, presenting an enhanced model, and evaluating the results via run-time compliance checking. Finally, we plan to employ the trust assessment to validate the model with more attributes and scenarios. 


\subsection*{Acknowledgements}
{\small The work was supported by ERDF "CyberSecurity, CyberCrime and Critical Information Infrastructures Center of Excellence" (No. CZ.02.1.01/0.0/0.0/16\_019/0000822) and ”Internal Grant Agency of Masaryk University” (Reg No. CZ.02.2.69/0.0/0.0/19\_073/0016943), registration number 1254/2021.}

\bibliographystyle{apalike}
{\small
\bibliography{example}}


\end{document}